%% file: main.tex
\documentclass[journal]{IEEEtran}
\input{sections/Preamble}
\begin{document}
\thispagestyle{fancy}
\bibliographystyle{IEEEtran}
\input{sections/title}
\input{sections/introduction}
\input{sections/preliminary}
\input{sections/motivation}

\input{sections/simulations}

\input{sections/Conclusions}
\input{main.bbl}

\newpage
\begin{center}
	\textbf{IEEE Copyright Notice}
\end{center}
\begin{ttfamily}
	© 2025 IEEE. Personal use of this material is permitted. Permission from IEEE must be obtained for all other uses, in any current or future media, including reprinting/republishing this material for advertising or promotional purposes, creating new collective works, for resale or redistribution to servers or lists, or reuse of any copyrighted component of this work in other works.
\end{ttfamily}

\begin{center}
	\textbf{Published Version}
\end{center}

This article has been published in IEEE Communications Letters. The final published version is available at \href{https://ieeexplore.ieee.org/abstract/document/10706909}{DOI:10.1109/LCOMM.2024.3475874}.
\end{document}

%% file: sections/Preamble.tex
\usepackage{amsmath,amsfonts,amssymb}
\usepackage{graphicx}
\usepackage{xcolor}
\usepackage{subfig}
\usepackage{makecell}
\usepackage{multirow}
\usepackage{tikz}
\usepackage{mathrsfs}
\usetikzlibrary{calc,arrows,shapes,chains,positioning,shapes.geometric}
\usepackage{enumitem}
\usepackage{algpseudocode}
\usepackage{algorithm}

\usepackage{amsthm,bm}
\usepackage{footnote}
\usepackage{pgfplots}
\pgfplotsset{width=10cm,compat=1.9}
\pgfplotsset{every axis legend/.style={
    cells={anchor=center},
    inner xsep=3pt,inner ysep=2pt,
    nodes={inner sep=2pt,text depth=0.1em},
    anchor=north east,
    shape=rectangle,
    fill=white,draw=green,font=\footnotesize,
}}
\usepackage{booktabs}
\usepackage{fancyhdr}
\usepackage{svg}

\fancypagestyle{plain}{
    \fancyhf{}
    \fancyfoot[C]{\thepage}
    
}

\usepackage{hyperref}
\hypersetup{
    anchorcolor=black, 
    bookmarksnumbered=true,
    bookmarksopen=true,
    bookmarksopenlevel=1,
    colorlinks=true,
    linkcolor=blue, 
    citecolor=red,
    urlcolor=magenta
}

\usepackage{titlesec}

\titlespacing*{\section}
{0pt}{*1}{*0.5}

\titlespacing*{\subsection}
{0pt}{*0.5}{*0.5}

%% file: sections/title.tex
\title{Boosting Ordered Statistics Decoding of Short LDPC Codes with Simple Neural Network Models}
\author{Guangwen Li, Xiao Yu
\thanks{G.Li is with  Shandong Technology and Business University, Yantai, China, lgwa@sdu.edu.cn}
\thanks{X.Yu is with  Binzhou Medical University, Yantai, China, YuXiao@bzmc.edu.cn}
}
\maketitle
\begin{abstract}
Ordered statistics decoding has been instrumental in addressing decoding failures that persist after normalized min-sum decoding in short low-density parity-check codes. Despite its benefits, the high computational complexity of effective ordered statistics decoding has limited its application in complexity-sensitive scenarios. To mitigate this issue, we propose a novel variant of the ordered statistics decoder. This approach begins with the design of a neural network model that refines the measurement of codeword bits, utilizing iterative information from normalized min-sum decoding failures. Subsequently, a fixed decoding path is established, comprising a sequence of blocks, each featuring a variety of test error patterns. The introduction of a sliding window-assisted neural model facilitates early termination of the ordered statistics decoding process along this path, aiming to balance performance and computational complexity. Comprehensive simulations and complexity analyses demonstrate that the proposed hybrid method matches state-of-the-art approaches across various metrics, particularly excelling in reducing latency.
\end{abstract}


%% file: sections/introduction.tex
\thispagestyle{fancy}
\section{Introduction}
\label{intro_sec}

Channel coding is essential for ensuring robust data transmission in telecommunication systems. Gallager’s low-density parity-check (LDPC) codes are renowned for their potential to approach the theoretical Shannon limit \cite{gallager62,mackay96}. However, in practical applications, the low-complexity normalized min-sum (NMS) variant \cite{chen2002near}, which serves as a substitute for belief propagation (BP) decoding, widens the gap to maximum likelihood (ML) performance, especially for short LDPC codes.

To bridge this performance gap, Nachmani et al. \cite{nachmani16,liang18} adapted iterative BP decoders into various neural network (NN) models, aiming to harness the strengths of NNs to enhance decoding. Buchberger et al. \cite{buchberger21} introduced a neural BP with decimation (NBP-D) approach, optimizing decoding through multiple choices. Kwak et al. \cite{kwak2023boosting} prepared another set of weights for the neural min-sum model to post-process the uncorrected sequences. Despite these efforts, a significant gap remains, making short LDPC codes less competitive than traditional codes like extended Bose–Chaudhuri–Hocquenghem (eBCH) codes in high-reliability sectors such as the Internet of Things (IoT) \cite{shirvanimoghaddam2018short}.

Ordered statistics decoding (OSD) \cite{Fossorier1995} has been extensively investigated to achieve ML performance for linear block codes with manageable computational complexity, aiming to reduce the average number of test error patterns (TEPs) required while maintaining frame error rate (FER) performance. Recent efforts on eBCH codes have applied threshold strategies to skip unlikely TEPs or to terminate decoding early, with validated effectiveness \cite{yue2021probability, choi2019fast, cavarec2020learning}. However, the adaptive ordering of TEPs in OSD decoding, determined on-the-fly with respect to received sequences, limits decoder throughput, making it challenging  to meet the stringent low-latency requirements of applications such as 6G \cite{yue2023efficient}.

Given the existence of parallelizable NMS or BP decodings for short LDPC codes, Baldi et al. \cite{baldi2015analysis,baldi2016use} proposed a hybrid framework where order-4 OSD is employed only when NMS decoding fails, supported by the ordering of TEPs in probability rather than conventional Hamming weight \cite{kabat2007new}, it was validated to achieve state-of-the-art (SOTA) FER, but its complexity and latency remain to be further reduced.
On the other hand, an architecture combining BP-adapted recurrent neural network (RNN) models, focused on the dominant absorbing sets of short LDPC codes, with order-2 OSD, has claimed near ML performance \cite{Rosseel2022}. However, this approach faces drawbacks such as intense hardware complexity and complicated training.

Building on the hybrid approach of NMS and OSD, the proposed scheme introduces three key distinctions:
- The magnitude of received log-likelihood ratios (LLRs) as reliability measurement is replaced.
- The process of generating an ordered TEP list is simplified.
- The hard threshold technique for early decoding termination is replaced.
 Specifically, the main contributions of this work are as follows:
\begin{itemize}
    \item A neural network model known as decoding information aggregation (DIA) is developed, which utilizes a posteriori LLRs from NMS decoding failures to create a novel measurement for each codeword bit.
    \item All TEPs are categorized into a sequence of blocks termed the decoding path, with ordering determined by the ratio of occurrences to block size, derived from the statistics of error patterns in NMS decoding failures.
    \item A sliding window-assisted (SWA) neural network model is integrated into OSD, targeting early termination to reduce complexity with minimal impact on performance.
\end{itemize}

The structure of this paper is as follows: Section \ref{preliminary} reviews the fundamentals of NMS and original OSD. Section \ref{motivation} delves into the motivations behind the decoding path, DIA, and SWA OSD. Section \ref{simulations} presents our experimental findings and complexity analysis. Finally, Section \ref{conclusions} concludes the paper.

%% file: sections/preliminary.tex
\section{NMS and Original OSD}
\label{preliminary}

Let $\mathbf{c}$ be a codeword of length $N$, encoded by a message vector $\mathbf{m}$ of length $K$ through the equation $\mathbf{c} = \mathbf{m}\mathbf{G}$ over the Galois Field GF(2), where $\mathbf{G}$ is the generator matrix. The $i$-th codeword bit $c_i$ is then modulated into a signal $s_i = 1 - 2c_i$ using the binary phase shift keying (BPSK) scheme. The received sequence at the additive white Gaussian noise (AWGN) channel output is given by $y_i = s_i + w_i$ for $i = 1,2,\cdots,N$, where $w_i$ is Gaussian distributed with zero mean and variance $\sigma^2$. The signal-to-noise ratio (SNR), expressed as $E_b/N_0$ in decibels, is calculated as $10 \log \left(\frac{N}{2K\sigma^2}\right)$. The magnitude of $y_i$ serves as a measure of reliability for the $i$-th hard decision $r_i = \mathbf{1}(y_i < 0)$, with $\mathbf{1}(\cdot)$ being the indicator function.

The NMS decoding aims to narrow the performance gap between BP and min-sum decoding with minimal additional complexity. It simplifies the update process at check nodes by replacing the traditional $\tanh$ or $\tanh^{-1}$ functions in BP with normalized min-sum terms. In its basic form, the shared normalization factor is optimized when the NMS is treated as a NN model. However, the presence of inherent short cycles in the Tanner graphs of short LDPC codes impedes NMS or BP decoders from achieving near ML performance. To overcome this, order-$p$ ordered OSD \cite{Fossorier1995} is initiated only after NMS decoding fails, a hybrid approach that has been proven effective in balancing performance and complexity \cite{baldi2015analysis,baldi2016use}.

Assuming the code's parity check matrix $\mathbf{H}$ of size $(N-K) \times N$ has full row rank, the conventional $\mathbf{H}$-oriented order-$p$ OSD begins by sorting the received sequence $\mathbf{y} = [y_i]_{i=1}^N$ in ascending order of magnitude to obtain $\mathbf{y}^{(1)}$, resulting in a column permutation $\pi_1(\mathbf{H}) = \mathbf{H}^{(1)}$. After Gaussian elimination, $\mathbf{H}^{(1)}$ is reduced to its systematic form $\mathbf{H}^{(2)} = \left[\mathbf{I} : \mathbf{Q}_2\right]$, with identity matrix $\mathbf{I}$ and matrix $\mathbf{Q}_2$ of dimensions $(N-K) \times (N-K)$ and $(N-K) \times K$, respectively. As a result, the entailed secondary column permutation $\pi_2(\cdot)$ leads to the swapping of $\mathbf{y}^{(1)}$ into $\mathbf{y}^{(2)}$. The $N-K$ leftmost codeword positions in $\mathbf{y}^{(2)}$ form the least reliable basis (LRB), while the remaining $K$ positions constitute the most reliable basis (MRB). Subsequently, for each codeword candidate $\overline{\mathbf{c}}_{j} = [\overline{\mathbf{c}}_{j1} : \overline{\mathbf{c}}_{j2}]$, $j=1,2,\cdots,\sum_{i=0}^{p}\binom{K}{i}$, $\overline{\mathbf{c}}_{j2}$ is evaluated as  the alternative bitwise sum of TEP  $\mathbf{e}_j$ whose Hamming weight is up to $p$ and hard decision of the bits indexed by the MRB, to derive $\overline{\mathbf{c}}_{j1} = \overline{\mathbf{c}}_{j2}\mathbf{Q}_2^\top$ under the parity check constraint $\mathbf{H}^{(2)} {\overline{\mathbf{c}}}_j^\top = \mathbf{0}$. The transmitted codeword $\hat{\mathbf{c}}$ is thus estimated as $\pi_1^{-1} \circ \pi_2^{-1}(\tilde{\mathbf{c}})$, where $\tilde{\mathbf{c}}$ is chosen to minimize the following criterion \eqref{argmin_dis} \cite{Fossorier1995}:
\begin{equation}
\label{argmin_dis}
\tilde{\mathbf{c}} = \arg \min_{\bar{\mathbf{c}}_j} \sum_{i = 1}^N \mathbf{1}(r_i^{(2)} \neq \overline{c}_{i}) |y_i^{(2)}|,
\end{equation}
with $\mathbf{r}^{(2)} = [r_i^{(2)}]_{i=1}^N$ being the hard decision of $\mathbf{y}^{(2)} = [y_i^{(2)}]_{i=1}^N$, and $\overline{c}_{i}$ denotes the $i$-th bit of $\overline{\mathbf{c}}_{j}$. The optimal weighted Hamming distance $h_k$ among the TEPs in the $k$-th TEP block, as defined by  \eqref{min_dis}, will be referenced later.
\begin{equation}
\label{min_dis}
h_k = \min_{\bar{\mathbf{c}}_j} \sum_{i = 1}^N \mathbf{1}(r_i^{(2)} \neq \overline{c}_{i}) |y_i^{(2)}|.
\end{equation}

%% file: sections/motivation.tex
\section{Decoding Path, DIA, and SWA OSD}
\label{motivation}
\begin{figure}[htbp]
    \centering
    \includegraphics[width=0.4\textwidth]{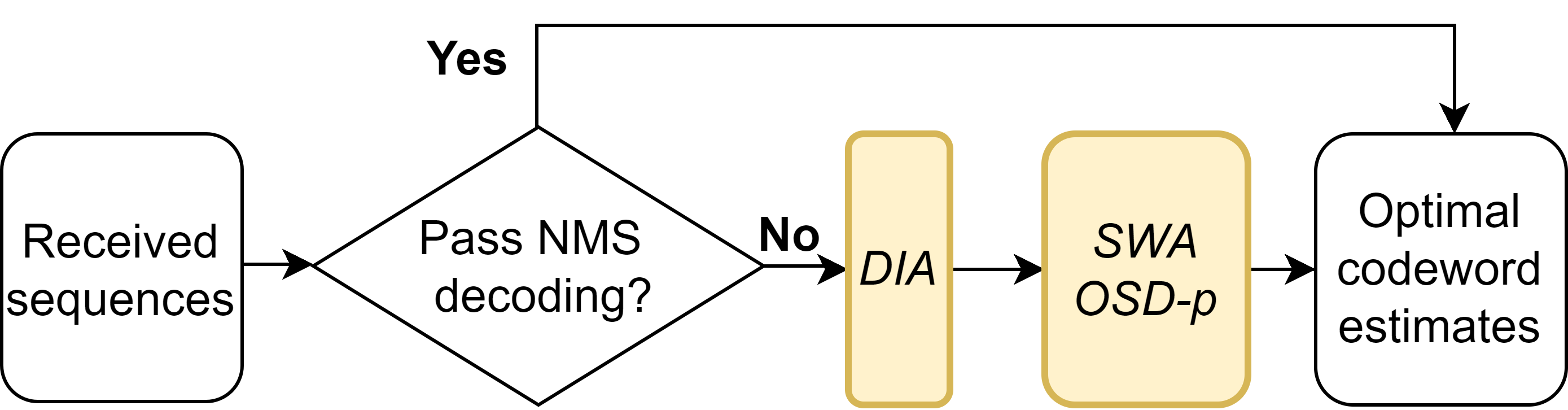}
    \caption{Flow chart of NMS-DIA-SWA OSD decoding}
    \label{architecture_decoding}
\end{figure}

The architecture depicted in Fig.~\ref{architecture_decoding} illustrates the synergistic interaction among the three components of the NMS-DIA-SWA OSD (N-D-S) decoding process. Initially, the NMS handles the majority of the decoding task, leveraging its advantages of low complexity and low latency, which significantly enhances the overall hybrid approach. Next, the iterative data from unsuccessful NMS decoding attempts are directed into the DIA model to refine the measurement of codeword bits. Finally, the OSD navigates through the TEPs organized within each block of the predefined decoding path, with the incorporated SWA model serving as an arbitrator to decide whether to terminate the decoding early. The following subsections delve into the innovative aspects of the highlighted blocks in Fig.~\ref{architecture_decoding}.

\subsection{Decoding Path Formation}
The practical application of order-$p$ OSD often encounters incomplete exploration of TEPs, making the scheduling of TEPs a critical factor in OSD performance \cite{baldi2016use}. This section presents a novel empirical approach to TEP ordering, based on the statistical distribution of actual error patterns, which replaces the method previously used in \cite{baldi2016use}. The process is divided into two stages: categorization and prioritization.

\textbf{Categorization:} The MRB range of $[1,K]$ for the code is evenly divided into $q$ segments (with the remainder merged into the last segment). Let $W_i$ denote the Hamming weight of the $i$-th segment. A permutation of the $q$ Hamming weights, uniquely identified by the tuple $\mathbf{W} = (W_1, W_2, \ldots, W_q)$, forms a TEP block to be filled with TEPs. The total number of TEP blocks is derived as $l_p = \binom{p+q}{q}$, subject to the constraint $\sum_{i=1}^{q} W_i \leq p$. Finally, all TEPs with Hamming weights up to $p$ are allocated to the respective TEP blocks if their segment weights align with $\mathbf{W}$.

\textbf{Prioritization:} For each received sequence $\mathbf{y}$ that leads to an NMS decoding failure, it is replaced by the DIA-enhanced measurement $\tilde{\mathbf{y}}$. The actual error pattern is then identified by making a hard decision on the MRB of the transformed sequence $\tilde{\mathbf{y}}^{(2)}$, which undergoes two column permutations specific to the OSD. As a result, the counter for the TEP block containing this error pattern is incremented. A decoding path is then established by analyzing a substantial dataset of failed NMS decoding instances to achieve statistical stability. This path consists of a sequence of TEP blocks $\bm{b}_i, i \in [1, l_p]$, with $l_p$ regarded as its nominal length, prioritized by the ratio of the counter value to the TEP block size (i.e., the number of enclosed TEPs) in descending order.

\begin{figure*}[htbp]
    \centering
    \subfloat[NN architecture of DIA Model\label{fig:dia}]{\includegraphics[width=0.65\linewidth]{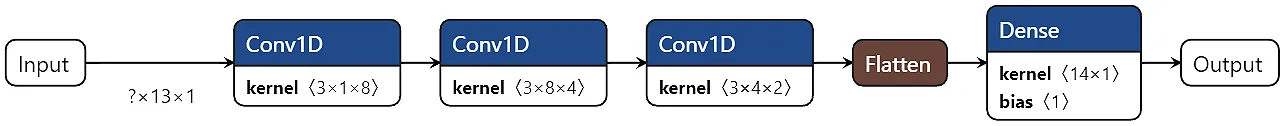}}
    \hfill
    \subfloat[NN architecture of SWA Model\label{fig:swa}]{\includegraphics[width=0.34\linewidth]{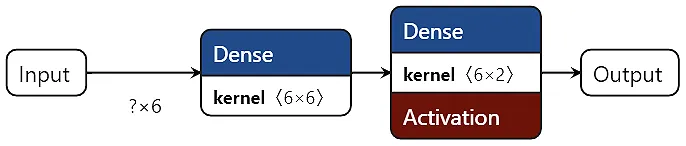}}
\caption{Input and output dimensions for the DIA model are '$? \times (T+1) \times 1$' and '$? \times 1$', respectively, while the SWA model has dimensions '$? \times (w_d + 1)$' and '$? \times 2$. Here, '?' denotes a variable batch size, with $T = 12$ and $w_d = 5$ for the LDPC (128, 64) code \cite{helmling19}.}

    \label{fig:dia_swa}
\end{figure*}
\subsection{DIA and SWA Models}
As shown in Fig.~\ref{fig:dia}, the DIA model is a four-layer Convolutional Neural Network (CNN) \cite{chollet2015keras}. It utilizes '3x1' filters across three 'Conv1D' layers to expand the receptive field, followed by a 'Dense' layer that produces a refined bit measurement. For each NMS decoding failure, $N$ training samples are generated simultaneously. Each sample is a combination of the received signal $y_i$ and the bit-wise trajectory of the a posteriori LLRs across iterations, with a total size of $T+1$, where $T$ is the maximum number of iterations. The label for each sample corresponds to the transmitted codeword bit.

In comparison, the shallow two-layer SWA model, depicted in Fig.~\ref{fig:swa}, outputs a binary probability distribution to determine whether to exit the current decoding process. The generation of SWA training samples is more intricate. Given the decoding path $\bm{b}_i$, where $i \in [1, l_p]$, the OSD transforms a new measurement $\tilde{\mathbf{y}}$ into $\tilde{\mathbf{y}}^{(2)}$ to derive $\bar{\mathbf{c}}_j$. Meanwhile, the received sequence $\mathbf{y}$ undergoes the same permutations, resulting in $\mathbf{y}^{(2)}$. The sequence $\mathbf{y}^{(2)}$ and its hard decision $\mathbf{r}^{(2)}$ are then substituted into the evaluation of \eqref{min_dis} for each $\bm{b}_i$, yielding a long list of $l_p$ positive numbers. From this list, a total of $l_p - w_d + 1$ training samples are generated by sliding a window of width $w_d$ with a unit stride. Each training sample consists of a sorted list of size $w_d + 1$, incorporating the window position $w_p$, which can empirically enhance model accuracy. If the ground truth weighted Hamming weight is included in the sample, it is labeled '0' to indicate decoding termination; otherwise, it is labeled '1'.

The DIA model employs a standard cross-entropy (CE) loss function for optimization, while the SWA model utilizes a weighted CE loss function defined as follows:

\begin{equation}
\label{ce_loss}
\ell_{\text{ce}} = \sum_{i=1}^{b_n} \alpha_i \beta_i \left( \sum_{j=0}^1 p(d_i = j) \cdot \log \frac{1}{p(\hat{d}_i = j)} \right)
\end{equation}

Here, $\alpha_i$, derived from the distribution of cases into binary classes, denotes the class weight for the $i$-th sample within a batch of size $b_n$ to address class imbalance. To mitigate adverse impacts on the FER, the penalty coefficient $\beta_i = \max(1, \gamma \cdot \mathbf{1}(d_i = 1, \hat{d}_i = 0))$, with $\gamma = 10$, imposes a biased penalty on predictions of decoding termination $\hat{d}_i = 0$ when ground truth $d_i = 1$ indicates that decoding should continue.

Given the simplicity of both models, with an initial learning rate of 0.001 that decays by a factor of 0.95 every 500 steps, training with the Adam optimizer \cite{kingma14} is expected to be completed within 20,000 steps or a few minutes on a desktop computer equipped with a 2.60 GHz Intel i7-6700HQ processor.

\subsection{SWA OSD}
\begin{algorithm}[ht]
\caption{\hspace{1cm}OSD Integrated with SWA Model}
\label{alg::sliding_osd}
\begin{algorithmic}[1]
\Require
Received sequence $\mathbf{y}$ and hard decision $\mathbf{r}$; $\tilde{\mathbf{y}} = [\tilde{y}_i]_{i=1}^N$ (DIA outputs); parity check matrix $\mathbf{H}$; decoding path $\bm{\mathit{b}}_i, i \in [1, l_p]$; window width $w_d$; soft margin $s_m$.
\Ensure
Optimal codeword estimate $\hat{\mathbf{c}}$ for $\mathbf{y}$.
\State
Using $\tilde{\mathbf{y}}$, perform the first two steps of the original OSD, resulting in $\tilde{\mathbf{y}}^{(2)} = \pi_2 \circ \pi_1(\tilde{\mathbf{y}})$ and $\tilde{\mathbf{H}}^{(2)} = \pi_2 \circ \pi_1(\mathbf{H}) = \left[ \mathbf{I} : Q_2 \right]$. Apply the same two permutations to obtain $\mathbf{y}^{(2)} = \pi_2 \circ \pi_1(\mathbf{y})$ and $\mathbf{r}^{(2)} = \pi_2 \circ \pi_1(\mathbf{r})$.
\State 
For $i = 1, 2, \dots, w_d$, compute $h_i$ for $\bm{b}_i$ using \eqref{min_dis}. Set the global minimum $g_m \gets \min\{h_i, i \in [1, w_d]\}$ and the window position $w_p \gets 0$.
\Repeat
\State Feed SWA model with sorted window elements $h_i, i \in [1, w_d]$ and $w_p$: output $[p_0, p_1] = [p(\hat{d} = 0), p(\hat{d} = 1)]$.
\If{$p_0 > s_m$} 
    {Break.}
\Else { Slide the window along the decoding path: $w_p \gets w_p + 1$, update $h_i, i \in [w_p, w_p + w_d]$ using \eqref{min_dis}.}
    \If{$h_{(w_p + w_d)} > g_m$} 
        {Go to Step 6.}
    \Else 
        { $g_m \gets h_{(w_p + w_d)}$.}
    \EndIf
\EndIf
\Until $w_p == l_p - w_d + 1$.
\State
\Return Retrieve the $\tilde{\mathbf{c}}$ associated with $g_m$, then $\hat{\mathbf{c}} \gets \pi_1^{-1} \circ \pi_2^{-1}(\tilde{\mathbf{c}})$.
\end{algorithmic}
\end{algorithm}

The inherent risk of confusing a local minimum with the global minimum during the design of early decoding termination necessitates the use of the SWA model to balance the trade-off between performance loss and complexity reduction through a tunable soft margin $s_m$. As outlined in Algorithm~\ref{alg::sliding_osd}, Steps 3 through 11 represent the core of this methodology, detailing the sliding window mechanism where conditional early termination is governed by the $s_m$ evaluation.

%% file: sections/simulations.tex
\section{Simulation Results and Analysis}
\label{simulations}
For LDPC CCSDS (128,64) code \cite{helmling19}, the normalization factor for NMS decoder with $T = 12$ iterations is optimized to be $0.78$. The DIA and SWA models were trained using NMS failures collected at SNR = 2.7dB, with $q = 6$ in partitioning the MRB. The hybrid architecture, illustrated in Fig.~\ref{architecture_decoding}, is referred to as N-D-S($p, \hat{l}_p, s_m$), where $\hat{l}_p$ denotes the length of the surviving TEP blocks when their corresponding nominal decoding path undergoes truncation or skipping.

PB-OSD \cite{yue2021probability} and FS-OSD \cite{choi2019fast} are SOTA OSD variants for eBCH codes, making it meaningful to evaluate their performance on LDPC codes as well. Specifically, PB$_0$ denotes the original settings from \cite{yue2021probability}, while both PB$_2$ (with \( \text{P}_t^{\text{pro}} \gets 0.1 \cdot \text{P}_t^{\text{pro}} \) and \( \text{P}_t^{\text{suc}} \gets \max(\text{P}_t^{\text{suc}}, 0.9) \)) and FS-OSD (with \( \tau_E = 6 \), \( \beta = 0.1 \), \( \tau_{PSC} = 30 \)) are adapted to achieve competitive FER performance.

All decoding schemes were implemented in Python and run on the TensorFlow platform using Google Colab. Interested readers can access the source code on GitHub\footnote{\url{https://github.com/lgw-frank/Short\_LDPC\_Decoding\_OSD}} to reproduce the decoding results presented in this paper.
\subsection{Impact of DIA on NMS Decoding Failures}

\begin{figure}[htbp]
    \centering
    \subfloat[CE evolution\label{cross_entropy_128}]{\includegraphics[width=0.48\linewidth]{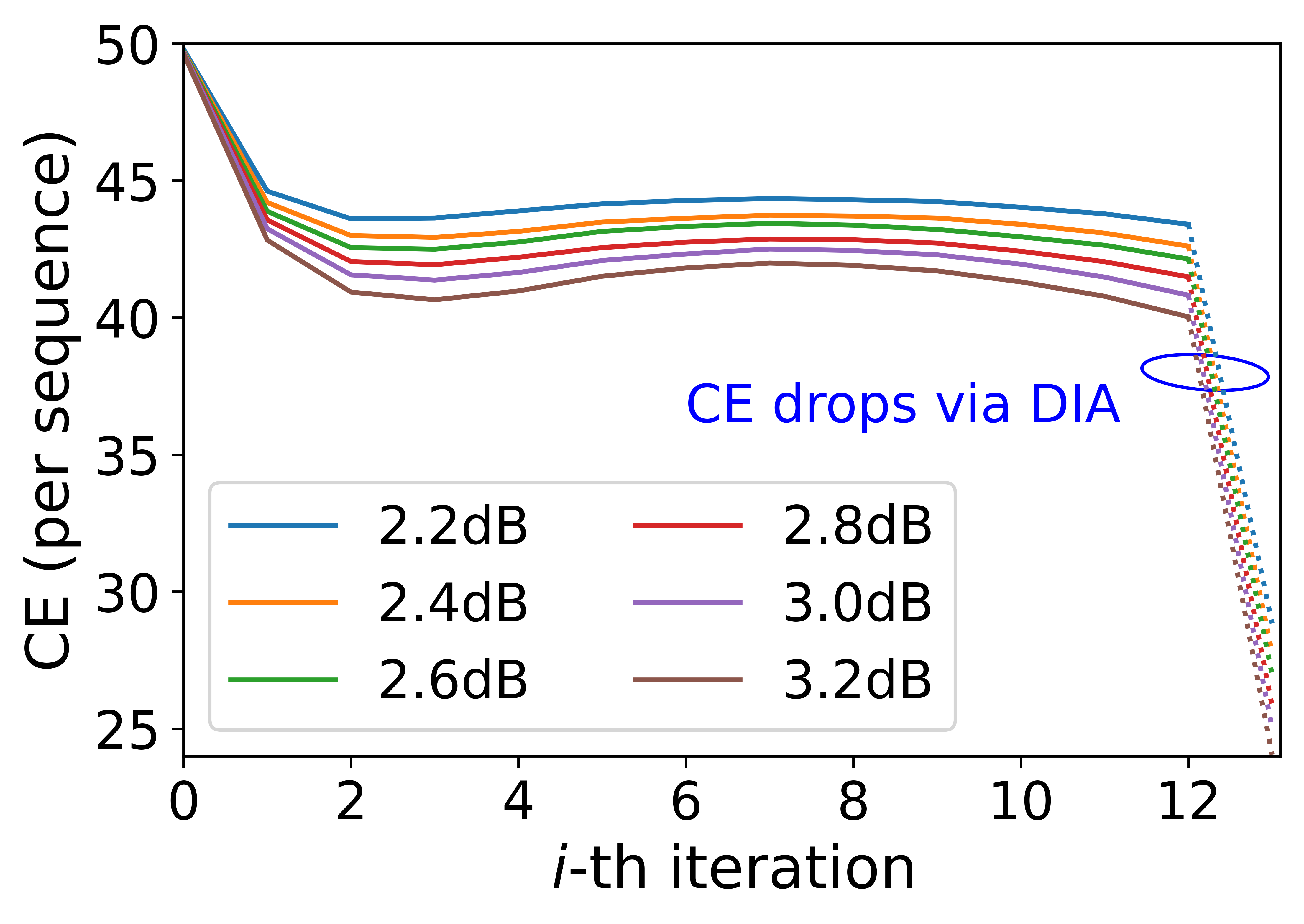}}
        \hfill 
    \subfloat[CDF of $\delta$\label{cdf128}]{\includegraphics[width=0.5\linewidth]{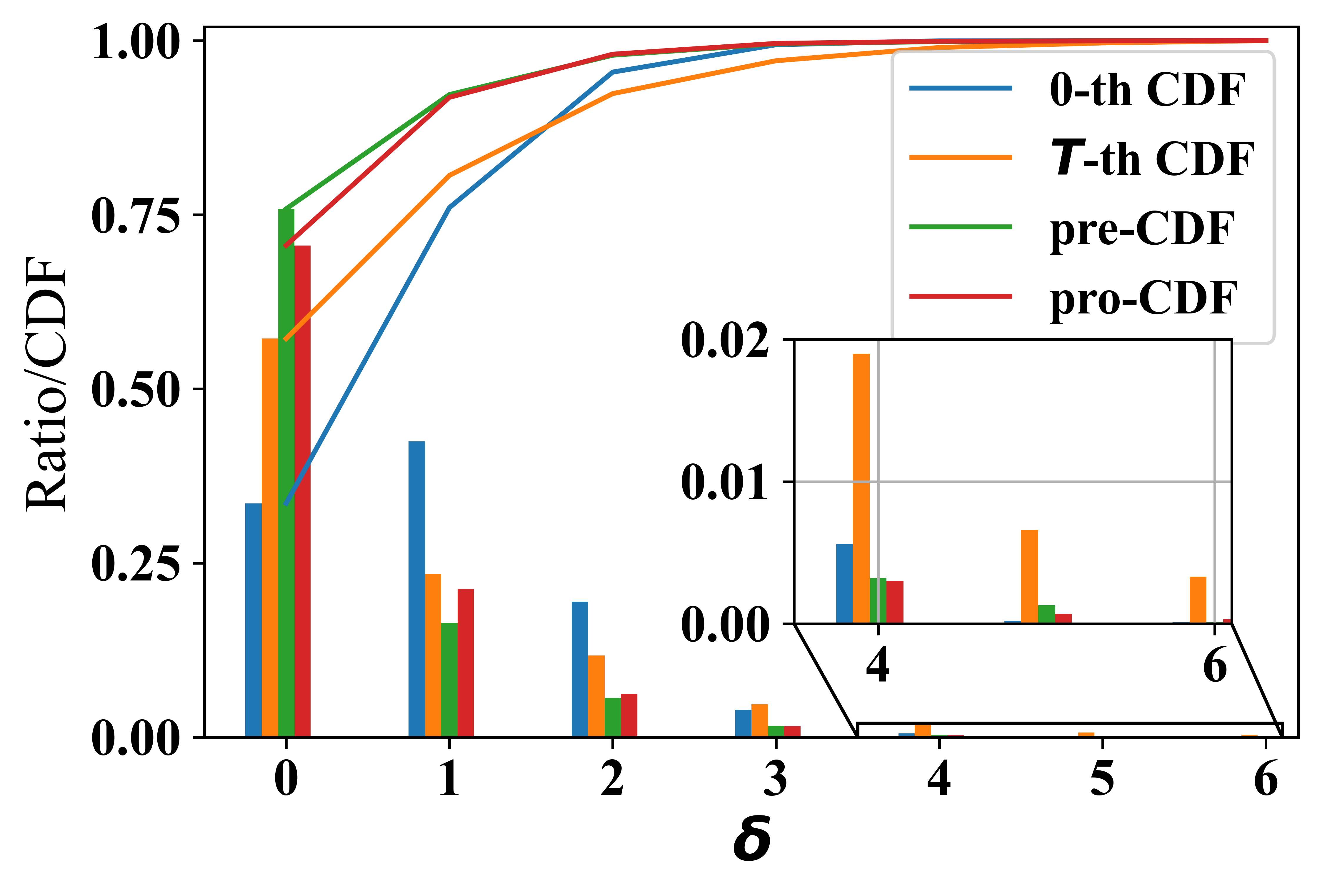}}
    \caption{CE of the $i$-th NMS iteration's a posteriori LLRs and derived DIA output from these LLRs; CDF of $\delta$ in the MRB of the LDPC (128,64) code (cases of $\delta \geq 6$ merged for clarity).}
    \label{cross_entropy__ber_plot128}
\end{figure}

For NMS decoding failures, the CE of a posteriori LLRs across each decoding iteration is compared to the DIA output, as shown in Fig.~\ref{cross_entropy_128}. The CE for each iteration decreases rapidly before stagnating, indicating that additional NMS iterations would not significantly improve the approach to the ground truth. However, the notable reduction in CE for the DIA output demonstrates its effectiveness in enhancing the magnitude of codeword bits with correct signs while suppressing others. This aligns with the objective of OSD to populate the MRB with the correct bits that have the largest magnitudes. Furthermore, the consistency of the CE curves across various SNRs suggests that a model trained at a specific SNR can generalize well across different SNR levels.

In Fig.~\ref{cdf128}, the cumulative distribution function (CDF) of the number of erroneous $\delta$ in the MRB is compared across various measurement sources for NMS decoding failures. The 'pre-CDF' and 'pro-CDF' curves, representing the DIA output before and after Gaussian elimination, respectively, consistently lie above the '0-th CDF' (initial received sequence) and the '$T$-th CDF' (final NMS iteration). Hence, the DIA model is capable of preparing more NMS decoding failures within the decodable scope of order-$p$ OSD. Notably, the theoretical bounds for the $\delta$ distribution, as discussed in \cite{dhakal2016error}, assume AWGN outputs and are thus too loose to be directly applied to these DIA outputs.

\subsection{Decoding Performance}
For short LDPC codes, NMS decoding is favored due to its low complexity and low latency. The effectiveness of combining NMS with the original OSD was extensively validated in \cite{baldi2016use}. We are therefore interested in examining whether the SOTA PB or FS variants can replace the original OSD and comparing them with the N-D-S scheme.

Regarding the same collection of NMS decoding failures, Fig.~\ref{fig:fer_n_at} presents a comparison among various decoders in terms of FER and the average number of TEPs $n_{at}$ required.

\begin{figure}[htbp]
    \centering
    \subfloat[FER of post-processing\label{fig:fer128_serial}]{
    \resizebox{0.49\linewidth}{!}{ 
    \input{plots/plot_128_fer_serial.tex}
    }
    }
    \hspace{-0.05\linewidth}  
    \subfloat[$n_{at}$ of post-processing\label{fig:fer128_pb_average_tep}]{
    \resizebox{0.48\linewidth}{!}{    \input{plots/average_128_teps.tex}
    }  
    }
    \caption{FER and $n_{at}$ comparison for post-processing NMS decoding failures of the LDPC (128,64) code.}
    \label{fig:fer_n_at}
\end{figure}

It is observed that PB$_2$ substantially outperforms PB$_0$ in terms of FER due to adapted thresholds, at the cost of a moderately increased $n_{at}$. The FS variant offers a slight FER advantage over PB$_2$, but with a much higher $n_{at}$. On the other hand, although the benchmarked N-D-S(3,30,1.0) scheme, prohibiting early decoding termination for its $s_m = 1$, is less attractive due to its $n_{at}$ evaluation, the N-D-S family rapidly excels in balancing FER and $n_{at}$ when $\hat{l}_{p}$ increases to 40 or 50 with a lowered $s_m = 0.8$.

Apart from evaluating the performance of component decoders individually, we are also interested in comparing the comprehensive FER of the proposed N-D-S scheme, defined as the product of the FER of NMS and SWA OSD, with popular decoders dedicated to LDPC codes.

\begin{figure}[htbp]
    \centering
    \input{plots/plot_128_fer}
    \caption{Comparison of various decoders for the LDPC (128,64) code.}
    \label{fig:fer128_whole}
\end{figure}

As shown in Fig.~\ref{fig:fer128_whole}, the fully parallelizable decoder NBP-D(10,4,4) \cite{buchberger21} significantly outperforms NMS($T$=12) and BP($T$=40) due to its incorporation of a learnable decimation procedure. However, due to inherent short cycles in the code structure, this NBP variant lags behind the N-D-S(3,60,0.9) and the D$_{10}$-OSD-2(25) \cite{Rosseel2022} by about 0.5 dB and 0.7 dB at FER $=10^{-4}$, respectively. Two configurations of another SOTA NMS+MRB(4) \cite{baldi2016use} with truncated sizes $10^4$ and $2 \cdot 10^5$ for the ordered TEP list, abbreviated as N-O(4,$e^4$) and N-O(4,$2e^5$), respectively, demonstrate competitive FERs in the high SNR region. N-O(4,$2e^5$) leads N-O(4,$e^4$) by about 0.5 dB, while the latter lags behind the proposed N-D-S(3,60,0.9) with a decoding path including 13,935 TEPs by about 0.2 dB. Notably, the N-O(4,$2e^5$) curve crosses the N-D-S(3,60,0.9) at SNR = 1.5 dB, indicating the role of DIA in correcting erroneous bits in the MRB at low SNR. The N-O(4,$2e^5$) configuration approaches its ML limit \cite{helmling19} within 0.06 dB and has a gap of 0.5 dB to the benchmarked sphere packing (SP) bound \cite{baldi2015analysis}.
\subsection{Complexity Analysis}

For the LDPC (128,64) code with average row and column weights $(d_c=8, d_v=4)$, the decoding complexity $C_{av}$ of a hybrid scheme can be evaluated as follows \cite{baldi2016use}:
\begin{equation}
    C_{av} = C_{NMS} + \xi \cdot C_{s}
    \label{weighted_complexity}
\end{equation}
where $\xi$ is approximated by the FER of NMS, and $C_{s}$ denotes the complexity associated with the specific OSD variant. Either time latency $c_t$ or computational complexity $c_f$, measured in floating-point operations (FLOPs), can be substituted into \eqref{weighted_complexity}.

For NMS, the complexity is approximately $2Nd_v + (N-K)(4d_c-3) = 2.88 \text{K}$ FLOPs per iteration, while for BP, NBP, or BP-RNN, it is $2Nd_v + (N-K)(3d_c-1) = 2.50 \text{K}$ FLOPs for simplicity. NBP-D(10,4,4) exhibits 18.2 times the complexity of NBP($T=50$) \cite{buchberger21}. D$_{10}$-OSD-2(25) can operate in either parallel or serial mode, concatenated by the order-2 OSD(s). The original $T=100$ was assigned for N-O(4,2$e^5$) \cite{baldi2015analysis}. For the referred code, half the size of its LRB is used to estimate the number of FLOPs for evaluating its weighted Hamming distance per TEP. Additionally, the DIA and SWA models require 167.7K and 0.06K FLOPs, respectively.

Table~\ref{tab:complexity-table} shows that the proposed N-D-O(3,60,0.9) significantly leads in terms of $c_t$ over the hybrids with SOTA components for eBCH codes. NBP-D(10,4,4) is less attractive due to its inferior FER and high $c_f$. Despite its superior FER, the intense complexity of 147.8K (parallel) or 82.8K (serial) makes D$_{10}$-OSD-2(25) less practical. The N-O(4,2$e^5$), which inspired our scheme, offers better FER than ours but requires inspecting significantly more TEPs on average. Considering the DIA's advantage revealed in Fig.~\ref{cdf128}, an N-O-S(4,$l_p$,$s_m$) with appropriately chosen $l_p$ and $s_m$ is expected to achieve a leading trade-off between performance and complexity. Particularly, when the order $p < 4$, aligning with the increasing CDF gap with decreased $p$ in Fig.~\ref{cdf128}, our scheme is more favorable. The last column of Table~\ref{tab:complexity-table} indicates our scheme has a smaller model size compared to the demanding NBP-D(10,4,4) or D$_{10}$-OSD-2(25) models. Despite the missing $c_t$ data for the last three decoders, it is inferred that they are less responsive than N-D-O(3,60,0.9) considering the measured $c_f$ and $n_{at}$.

\begin{table}[htbp]
\caption{Complexity Comparison of Various Decoders for LDPC CCSDS (128,64) Code \cite{helmling19} at SNR = 3.5 dB}
\label{tab:complexity-table}
\resizebox{0.45\textwidth}{!}
{%
\begin{tabular}{|c|cccc|}
\hline
\multirow{2}{*}{Decoding Scheme} & \multicolumn{4}{c|}{Metrics}                                                                     \\ \cline{2-5} 
 & \multicolumn{1}{c|}{\begin{tabular}[c]{@{}c@{}}$c_t$\\ (ms)\end{tabular}} &
 \multicolumn{1}{c|}{\begin{tabular}[c]{@{}c@{}}$c_f$\\ (K)\end{tabular}} &
 \multicolumn{1}{c|}{\begin{tabular}[c]{@{}c@{}}$n_{at}$ \\ of OSD\end{tabular}} &
 \begin{tabular}[c]{@{}c@{}}Number \\ of Weights\end{tabular} \\ \hline
NMS-FS($p$=3)                     & \multicolumn{1}{c|}{152.5} & \multicolumn{1}{c|}{61.8}       & \multicolumn{1}{c|}{1250} & 1     \\ \hline
NMS-PB$_2$($p$=3)                 & \multicolumn{1}{c|}{95.7}  & \multicolumn{1}{c|}{10.3}  & \multicolumn{1}{c|}{150}  & 1     \\ \hline
N-D-O(3,60,0.9)                   & \multicolumn{1}{c|}{31}    & \multicolumn{1}{c|}{14.5}       & \multicolumn{1}{c|}{175}  & 160   \\ \hline
NBP-D(10,4,4) \cite{buchberger21} & \multicolumn{1}{c|}{-}     & \multicolumn{1}{c|}{45.5}       & \multicolumn{1}{c|}{0}    & 1553  \\ \hline
D$_{10}$-OSD-2(25) \cite{Rosseel2022} & \multicolumn{1}{c|}{-}     & \multicolumn{1}{c|}{147.8/82.8} & \multicolumn{1}{c|}{2081} & 10240 \\ \hline
N-O($p$=4,2$e^5$) \cite{baldi2016use} & \multicolumn{1}{c|}{-}     & \multicolumn{1}{c|}{33.4K}      & \multicolumn{1}{c|}{2528} & 1     \\ \hline
\end{tabular}
}
\end{table}

%% file: plots/plot_128_fer_serial.tex
	\begin{tikzpicture}[scale=0.55]
		\begin{semilogyaxis}[
			scale = 0.75,
			xlabel={$E_b/N_0$(dB)},
			ylabel={FER},
			xmin=2.2, xmax=3.2,
			ymin=4.5e-3, ymax=1.1e-1,
			xtick={2,2.2,...,3.4},
			legend pos = north east,
			ymajorgrids=true,
			xmajorgrids=true,
			grid style=dashed,
			legend style={legend columns=2},
                xminorgrids=false, 
                yminorgrids=true,
			]
\addplot[
color=black,
mark=halfcircle,
very thin
]
coordinates {
(2.0, 0.05219)
(2.2, 0.0384) 
(2.4, 0.02606)
(2.6, 0.01947)
(2.8, 0.01422) 
(3.0, 0.01075) 
(3.2,0.00732)
};
\addlegendentry{N-D-S(3,30,1.0)}

\addplot[
color=blue,
mark= square,
very thin
]
coordinates {
(2.0, 0.04827)
(2.2, 0.03604) 
(2.4, 0.02699)
(2.6, 0.01862)
(2.8, 0.01403)
(3.0, 0.01094)
(3.2, 0.00795)
};	
\addlegendentry{N-D-S(3,40,0.8)}
\addplot[
color=violet,
mark=triangle,
very thin
]
coordinates {
(2.0,0.03627)
(2.2,0.02566)
(2.4,0.01825)
(2.6,0.01329)
(2.8,0.00966)
(3.0,0.00687)
(3.2,0.00473)
};	
\addlegendentry{N-D-S(3,50,0.8)}

\addplot[
color=purple,
mark=*,
very thin
]
coordinates {
(2.2, 0.0463)
(2.4, 0.0431)
(2.6, 0.0412)
(2.8, 0.0379)
(3.0, 0.0336)
(3.2, 0.0304)
};
\addlegendentry{PB$_0(p=3)$\cite{yue2021probability}}


\addplot[
color=red,
mark=triangle*,
very thin
]
coordinates {
(2.0,0.0404)
(2.2, 0.0295)
(2.4, 0.0235)
(2.6, 0.0183)
(2.8, 0.0147)
(3.0, 0.0121)
(3.2, 0.0095)
};	
\addlegendentry{PB$_2(p=3)$\cite{yue2021probability}}
\addplot[
color=cyan,
mark=square*,
very thin
]
coordinates {
(2.0,0.0387)
(2.2, 0.0307)
(2.4, 0.0236)
(2.6, 0.0176)
(2.8, 0.0137)
(3.0, 0.0112)
(3.2, 0.0084)
};	
\addlegendentry{FS($p=3$)\cite{choi2019fast}}
  \end{semilogyaxis}
	\end{tikzpicture}

%% file: plots/average_128_teps.tex
	\begin{tikzpicture}[scale=0.55]
		\begin{semilogyaxis}[
			scale = 0.75,
			xlabel={$E_b/N_0$(dB)},
			ylabel={$n_{at}$},
			xmin=2.2, xmax=3.2,
			ymin=6e1, ymax=2e4,
			xtick={2,2.2,...,3.5},
			legend pos = north east,
			ymajorgrids=true,
			xmajorgrids=true,
			grid style=dashed,
			legend style={legend columns=2},
                xminorgrids=false, 
                yminorgrids=true,
			]
\addplot[
color=black,
mark=halfcircle,
very thin
]
coordinates {
(2.0, 1898.0 )
(2.2, 1898.0 )
(2.4, 1898.0 )
(2.6, 1898.0 )
(2.8, 1898.0 )
(3.0, 1898.0 )
(3.2, 1898.0 )
};
\addlegendentry{N-D-S(3,30,1.0)}
\addplot[
color=blue,
mark= square,
very thin
]
coordinates {
(2.0, 359)
(2.2, 270)
(2.4, 203)
(2.6, 158)
(2.8, 121)
(3.0, 92)
(3.2, 70)
};	
\addlegendentry{N-D-S(3,40,0.8)}
\addplot[
color=violet,
mark=triangle,
very thin
]
coordinates {
(2.0, 947)
(2.2, 680)
(2.4, 499)
(2.6, 372)
(2.8, 281)
(3.0, 189)
(3.2, 137)
};	
\addlegendentry{N-D-S(3,50,0.8)}

\addplot[
color=purple,
mark=*,
very thin
]
coordinates {
(2.2, 520.40)
(2.4, 403.64)
(2.6, 284.33)
(2.8, 201.26)
(3.0, 145.98)
(3.2, 108.92)
};
\addlegendentry{PB$_0(p=3)$\cite{yue2021probability}}
\addplot[
color=red,
mark=triangle*,
very thin
]
coordinates {
(2.0, 1524.33)
(2.2, 1092.50)
(2.4, 818.72)
(2.6, 559.43)
(2.8, 421.08 )
(3.0, 300.95)
(3.2, 220.19)
};	
\addlegendentry{PB$_2(p=3)$\cite{yue2021probability}}
\addplot[
color=cyan,
mark=square*,
very thin
]
coordinates {
(2.0, 4870.37)
(2.2, 3983.60)
(2.4, 3384.76)
(2.6, 2693.59)
(2.8, 2343.43 )
(3.0, 1929.84)
(3.2, 1530.99)
};	
\addlegendentry{FS($p=3$)\cite{choi2019fast}}
\end{semilogyaxis}
\end{tikzpicture}
	

%% file: plots/plot_128_fer.tex
\begin{tikzpicture}	[scale=0.55]
\begin{semilogyaxis}[
                scale = 1.,
			xlabel={$E_b/N_0$(dB)},
			ylabel={FER},
			xmin=0.0, xmax=4.0,
			ymin=1e-5, ymax=1.,
			xtick={0.0,0.5,1.0,...,4.0},
			legend pos = south west,
			ymajorgrids=true,
			xmajorgrids=true,
                grid style={dashed, thin},
                xminorgrids=false, 
                yminorgrids=true,
                width=0.7\textwidth, 
                height=0.5\textwidth, 
			]
\addplot[
color=orange,
mark=x,
solid,
very thin
]
coordinates {
(0.0, 0.96637) 
(0.5, 0.91191) 
(1.0, 0.80392) 
(1.5, 0.64575) 
(2.0, 0.43729)
(2.2,0.3698)
(2.4,0.2929)
(2.6,0.2203)
(2.8,0.1556)
(3.0,0.1077)
(3.2,0.0618)
(3.4,0.03771)
(3.6,0.02181)
(3.8, 0.01191)
(4.0, 0.0062)
};	
\addlegendentry{NMS($T$=12)}

\addplot[
color=purple,
mark= +,
solid,
very thin
]
coordinates {
(0.0, 0.955) 
(0.5, 0.905) 
(1.0, 0.80) 
(1.5, 0.7) 
(2.0,0.38)
(2.5,0.18)
(3.0,0.061)
(3.5,0.012)
(4.0,0.0023)
};	
\phantomsection
\addlegendentry{BP($T$=40)\cite{helmling19}}
\addplot[
color=magenta,
mark=halfcircle,
very thin
]
coordinates {
(1.0,2.3e-1)
(1.5,1.08e-1)
(2.0,5e-2)
(2.5,1.5e-2)
(3.0,4e-3)
(3.5,8e-4)
(4.0,1.2e-4)
};
\addlegendentry{\texorpdfstring{ NBP-D(10,4,4) \cite{buchberger21}}{NBP-D(10,4,4) \cite{buchberger21}}}
\addplot[
color=red,
mark=triangle,
very thin
]
coordinates {
(0.0,7.2e-1)
(0.5,3.6e-1)
(1.00, 1.7e-01)
(1.50, 7.e-02)
(2.00, 2.0e-02)
(2.50, 5.e-03)
(3.00, 1.05e-03)
(3.50, 1.5e-04)
(4.0,2.e-5)
};	
\phantomsection
\addlegendentry{N-O(4,$e^4$)\cite{baldi2016use}} 
\addplot[
color=blue,
mark= diamond,
]
coordinates {
(0.0,0.450676)
(0.5,0.292951)
(1.0, 0.139078)
(1.5,0.050272)
(2.0, 0.013954)
(2.2, 0.008561)
(2.4, 0.004484)
(2.6, 0.002326)
(2.8, 0.001193)
(3.0, 0.000569)
(3.2,0.000249)
(3.4, 0.000118)
(3.6,5.9e-05)
(3.8,2.7e-05)
(4.0,1.1e-05)
};
\addlegendentry{N-D-S-(3,60,0.9)}
\addplot[
color=cyan,
mark=triangle,
very thin
]
coordinates {
  (2.50, 2.2e-03)
  (2.75, 8.0e-4)
  (3.00, 3.0e-04)
  (3.22, 1.0e-04)
};	
\phantomsection
\addlegendentry{D$_{10}$-OSD-2(25) \cite{Rosseel2022}}

\addplot[
color=violet,
mark=triangle*,
very thin
]
coordinates {
(0.0,7e-1)
(0.5,3.3e-1)
(1.00, 1.5e-01)
(1.50, 5.0e-02)
(2.00, 1.0e-02)
(2.50, 1.3e-03)
(3.00, 1.5e-04)
(3.50, 1.3e-05)
(4.0,9e-7)
};	
\phantomsection
\addlegendentry{N-O(4,$2e^5$)\cite{baldi2016use}}

\addplot[
color=darkgray,
mark=square,
very thin
]
coordinates {
  (1.00, 1.064e-01)
  (1.50, 3.397e-02)
  (2.00, 8.773e-03)
  (2.50, 1.168e-03)
  (3.00, 1.321e-04)
  (3.50, 1.022e-05)
};	
\phantomsection
\addlegendentry{Simulated ML \cite{helmling19}}
\addplot[
color=black,
mark= square*,
very thin
]
coordinates {
  (0.00, 3.8e-01)
  (0.50, 1.9e-01)
  (1.00, 7e-02)
  (1.50, 1.9e-02)
  (2.00, 3.1e-03)
  (2.50, 3e-04)
  (3.0, 1.6e-5)
  (3.5,4.3e-7)
};	
\phantomsection
\addlegendentry{SP59 \cite{baldi2015analysis}}

\end{semilogyaxis}
\end{tikzpicture}

%% file: sections/conclusions.tex
\section{Conclusions}
\label{conclusions}
We introduced a hybrid decoder that incorporates three key innovations to enhance the post-processing performance of the component OSD. These innovations include: a neural network model designed to improve the reliability measurement of codeword bits from the failed iterative trace of normalized min-sum decoding; an empirical method for hierarchically sorting TEPs to establish a universally applicable decoding path; and a neural network model that facilitates early termination of OSD through real-time evaluation of elements in a sliding window along the decoding path.

The proposed hybrid preserves all the benefits of the original NMS+OSD framework, such as competitive FER, low complexity, low latency, and independence from noise estimation. Furthermore, with a fixed parameter configuration, it demonstrates strong resilience across various decoding scenarios within the $0 \, \sim 4 \, \text{dB}$ SNR range.